\documentclass{aa}  
\usepackage{graphicx}
\usepackage{indentfirst}
\usepackage{textcomp}
\usepackage{color}
\usepackage{amsmath}
\usepackage{amssymb}
\usepackage{txfonts}
\usepackage{mymacros}
\definecolor{darkgreen}{rgb}{0,0.4,0}

\begin{document} 

\authorrunning{Prabhu et al.}

   \title{Helicity proxies from linear polarisation of solar active regions}
   \author{A. Prabhu
          \inst{1},
          A. Brandenburg\inst{2,3,4,5},
          M. J. K\"apyl\"a\inst{6,1,2},
          \and
          A. Lagg\inst{1}
          }

   \institute{Max Planck Institute for Solar System Research, Justus-von-Liebig-Weg 3, D-37077 G\"ottingen, Germany\\
              \email{prabhu@mps.mpg.de}
          \and 
          NORDITA, KTH Royal Institute of Technology and Stockholm University, Roslagstullsbacken 23, SE-10691 Stockholm, Sweden
          \and
          Department of Astronomy, AlbaNova University Center, Stockholm University, SE-10691 Stockholm, Sweden
          \and
          JILA and Laboratory for Atmospheric and Space Physics, University of Colorado, Boulder, CO 80303, USA
          \and
          McWilliams Center for Cosmology \& Department of Physics, Carnegie Mellon University, Pittsburgh, PA 15213, USA
         \and
         Department of Computer Science, Aalto University, PO Box 15400, FI-00076 Aalto, Finland
             }

   \date{Received 29 January, 2020/ Accepted 16 June, 2020}


    \abstract
   {The $\alpha$ effect is believed to play a key role in the generation
   of the solar magnetic field. A fundamental test for its significance in
   the solar dynamo is to look for magnetic helicity of opposite signs 
  both between the two hemispheres as well as between small and large scales. However, measuring
  magnetic helicity is compromised by the inability to fully infer the
  magnetic field vector from observations of solar spectra, 
  caused by what is known as the $\pi$ ambiguity of 
  spectropolarimetric observations.}
   {We decompose linear polarisation into parity-even and parity-odd $E$ and $B$ polarisations, 
   which are not affected by the $\pi$ ambiguity.
   Furthermore, we study whether the correlations of spatial Fourier
   spectra of $B$ and parity-even quantities such as $E$ or
   temperature $T$ are a robust proxy for magnetic helicity of solar magnetic fields. }
  {We analysed polarisation measurements of active regions observed by the
  Helioseismic and Magnetic Imager on board the Solar Dynamics observatory. Theory predicts
  the magnetic helicity of active regions to have, statistically, opposite signs in the two hemispheres. 
   We then computed the parity-odd $EB$ and $TB$ correlations and tested for a systematic preference of
   their sign based on the hemisphere of the active regions. }
   {We find that: (i) $EB$ and $TB$ correlations are a reliable proxy for magnetic helicity, when 
   computed from linear polarisation measurements away from spectral line cores; and (ii)
  $E$ polarisation reverses its sign close to the line core. Our analysis reveals that Faraday
  rotation does not have a significant influence on the computed parity-odd correlations.}
   {The $EB$ decomposition of linear polarisation appears to be a good proxy for magnetic helicity
   independent of the $\pi$ ambiguity. This allows us to routinely infer magnetic helicity
   directly from polarisation measurements.}
    {}
   \maketitle

\section{Introduction}\label{intro}

Astrophysical bodies such as stars, galaxies, and planets are
known to posses magnetic fields,
typically on scales as large as those systems themselves.
Dynamo theory aims to describe the mechanisms responsible for
the generation and maintenance of these magnetic fields from first
principles.
Specifically, the solar magnetic field and its spatio-temporal
features (such as the cyclic polarity reversals) are ascribed to a dynamo
process acting within the Sun's convection zone.
One scenario attempts to explain the 
origin of solar magnetism with the $\alpha$ effect
\cite[]{Moffat_1978,Krause_Radler,BSS_12}.
In this framework, kinetic helicity (a measure of handedness)
of the gas motions, is believed to play a central role in the
generation of large-scale magnetic fields in the Sun.
This also results in the production of magnetic helicity, which
can be interpreted in terms of twist of flux tubes
or linkage of magnetic field lines \cite[]{Berger_Field,Blackman}.
The volume integral of the magnetic helicity density is almost perfectly
conserved---even in non-ideal magnetohydrodynamics (MHD);
see \cite{Berger84}.
This imposes important constraints on the evolution of magnetic
fields via a dynamo mechanism \citep{Bran+sub_2005}.
For the solar dynamo, the combined effect of stratification and
global rotation are believed to be responsible for the $\alpha$ effect
\citep{Krause_Radler}.
The $\alpha$ effect encapsulates the helical nature of turbulence within
the solar convection zone.
A key consequence of the $\alpha$ effect is the presence of opposite
signs
of magnetic helicity at small and large scales
\citep{Seehafer}.
Such magnetic fields are now referred to as bihelical \citep{YB03}
or doubly helical \citep{BB03}.
Additionally, due to the reflectional antisymmetry of cyclonic convection
across the equator, $\alpha$ changes sign at the equator \citep{Par55}.
Consequently, the magnetic helicity not only has opposite signs at large
and small scales, but it also changes sign across the equator.
Thus, from theoretical considerations we expect a hemispheric sign rule
for magnetic helicity in the Sun.
Specifically, one expects a positive (negative) sign of magnetic helicity
at large (small) scales in the Northern hemisphere and vice versa in
the Southern hemisphere.
Here, a small-scale field is defined as the difference between
actual and averaged fields.
In that sense, even the scale of active regions (ARs) must be regarded as
`small' because the large-scale field as seen in the solar butterfly diagram,
is obtained through azimuthal averaging, which also washes out ARs.

Much effort has been devoted to characterising the behaviour of magnetic
helicity in the Sun.
The primary motivation is to test the predictions of the $\alpha$
effect and thus indirectly verify the significance of the $\alpha$ effect
for the solar dynamo.
The earliest efforts were those of \cite{Seehafer90} and \cite{Pevtsov1995},
who analysed the magnetic field in local Cartesian patches and used the
vertical or $z$ component of the current helicity
$\left<j_z b_z\right>$ as a proxy for magnetic helicity.
Here $\vec{j}\propto\vec{\nabla}\times\vec{b}$ is the current density
and $\vec{b}$ is the magnetic field.
These studies focused on the helicity associated with ARs,
and they found it to be mostly positive (negative) in the Southern
(Northern) hemisphere.
However, given the aforementioned dependence of helicity on the scale, a more complete
picture can be obtained by looking at the spectra of magnetic helicity
\citep{Zhang_2014,Zhang_2016}.
A more global approach, taking into account the change in sign of
helicity across the equator was developed by \cite{Bran_2017}, called
the two-scale approach after \cite{Roberts_soward}.
This was followed by a systematic study employing this two-scale
approach over a large sample of Carrington rotations from solar cycle 24 by \cite{Singh18}.
They provided evidence for the expected hemispheric sign rule in the Sun,
specifically during the rising phase of cycle 24.

All the studies mentioned so far rely on the determination of the magnetic 
field on the Sun's photosphere. This is usually done by measuring
the full Stokes vector, ($I,Q,U,V$), where $I$ is the intensity, $Q$ and $U$
are the components of linear polarisation, and $V$ is circular polarisation.
Typically, for the
retrieval of the magnetic field at the photosphere, the Zeeman effect is 
used as a diagnostic. One attempts to 
deduce an atmospheric stratification that best fits the spectropolarimetric 
observations \citep{delToro}. Thus, the magnetic field vector is not a direct 
measurement but rather an inference. In addition, the use of Zeeman diagnostics 
bears an intrinsic ambiguity, referred to as the $\pi$ ambiguity,
associated with the transverse (perpendicular to the line-of-sight)  
component of the magnetic field. That is, we can only see it as an arrow-less vector
in the line-of-sight coordinate system. For the conversion to a solar
coordinate system, several disambiguation methods exist, based on
potential field extrapolations or on minimum energy techniques; see 
\cite{Metcalf} for a review.
However, these methods have limitations and fail to work accurately
in complex magnetic field topologies or where the determination of the
field is strongly influenced by the noise in the measurement.
The errors introduced by these disambiguation methods can have a
drastic impact on the computation of magnetic helicity. Hence, a means of
inferring the helicity of magnetic fields, independent of the $\pi$ 
ambiguity, is desired.

\cite{Bran_eb_1} introduced a possible proxy for helical magnetic fields, 
which could circumvent the uncertainty introduced by the $\pi$ ambiguity. 
They used Stokes $Q$ and $U$ polarisation measurements, and decomposed them 
into rotationally invariant $E$ and $B$ polarisations \citep{Kamionkowski_1997,Seljak_1997,Durrer}.
The $E$ and $B$ polarisations are parity-even and parity-odd quantities, respectively.
Correlations of $B$ polarisation with parity-even quantities such as $E$ 
polarisation or temperature $T$ can be indicative of the helicity of the 
underlying magnetic field \citep[]{Pogosian,Kahniashvili2005,Kahniashvili2014}. We expect that the sign of 
magnetic helicity changes across the equator at both large and small length scales. 
Thus, we expect the $EB$ correlation to reflect this behaviour and have 
systematically different signs in the two hemispheres. 
\cite{Bran_eb_1} used 
this $EB$ decomposition and tested it with full disk polarisation data from the
Vector SpectroMagnetograph (VSM) instrument of the Synoptic
Optical Long-term Investigations of the Sun (SOLIS) project. However,
they did not find significant non-zero parity-odd correlations from their 
analysis. \cite{Bran_eb_2} extended this work to a
fully global approach using spin-weighted spherical harmonics.
He focussed on the calculation of a global spectrum of the $EB$
correlation by taking into account its systematic sign change across
the equator.
Local aspects and features of the $E$ and $B$ patterns were completely
ignored, however.

For the present analysis, we adopt the local approach and focus on linear 
polarisation measurements of ARs from both hemispheres. We use the 
polarisation measurements obtained by the {\em Solar Dynamics Observatory} 
({\em SDO}). We then decompose this linear 
polarisation into $E$ and $B$ polarisations. The aim of this study is
to test if there are significant non-vanishing $EB$ correlations
from solar ARs and if they show a systematic preference
of a sign based on hemisphere. Therefore, we look at a sample of ARs and
analyse the $EB$ correlations and patterns computed from them in detail.
However, there are a few 
drawbacks of using polarisation data as is. They may have some systematic 
instrumental effects that need to be accounted for. Additionally, there are 
Doppler shifts of spectral lines and magneto-optical effects which also leave 
their imprint on the measured spectropolarimetric observations. 
Due to Faraday rotation, which is one of two magneto-optical effects, a constant
non-helical magnetic field can give rise to a non-zero $EB$ correlation.
This property was utilised in
theoretical studies of the cosmic microwave background
radiation \citep{Kosowsky+Loeb96,Scannapieco,Scoccola}.
However, for the purpose of this study, it is necessary to disentangle the contributions of the
intrinsic helicity of magnetic fields from those of Faraday rotation,
because both can cause a non-zero $EB$ correlation.

In Sect.~\ref{theory}, we briefly review the  motivations for the $EB$ 
decomposition and its relation to linear polarisation. In Sect.~\ref{data},
we discuss the observations and define correlation
spectra that we determine from those observations.
We also address the influence of 
Faraday rotation on our conclusions.
We conclude with a discussion and interpretation of our results in Sect.~\ref{dis}.

\section{$E$ and $B$ polarisations} \label{theory}

We begin by recalling some basics of polarised radiative transfer.
Let $\vec{I}(\tau_c) = (I,Q,U,V)^T$ be the Stokes vector for which the
radiative transfer equation (RTE) can be written as
\begin{equation}
	\frac{\dd \vec{I} }{\dd \tau_c} = \vec{K} (\vec{I} - \vec{S}).\label{rte}
\end{equation}
Here $\tau_c$ is the optical depth at the continuum wavelength, and $\vec{K}$ is
the propagation matrix, wherein the diagonal terms correspond to absorption,
and the off-diagonal terms are responsible for dichroism and dispersion.
The latter exchanges the states of polarisation caused by phase shifts during the
propagation, which includes the following two magneto-optical effects:
the exchange between the linear polarised components ($Q$ and $U$) is called
Faraday rotation, and between linear and circular polarised components ($Q$, $U$ and $V$) Faraday
pulsation.
$\vec{S}$ is the source-function
vector, which, under the assumption of local thermodynamic equilibrium (LTE),
can be approximated as $\vec{S} \equiv (B_\nu(T),0,0,0)$, where
$B_\nu(T)$ is the Planck function.
The measured quantity is $\vec{I}
(\tau_c = 0)$, and it is given by the formal solution of the RTE
\citep{Landi}. The observable complex linear
polarisation, $P(x,y) = Q+iU$, can be decomposed into the rotationally invariant
parity-even and parity-odd $E$ and $B$ polarisations, respectively.
Here, $x$ and $y$ are local Cartesian coordinates on the solar disk.
We thus invoke the small-scale limit, that is, we focus on small patches
on a sphere, where the curvature can be neglected.
The amplitudes of Stokes $Q$ and $U$ depend on the orientation of the 
polarisation basis.
It is thus desirable to transform this linear polarisation into quantities
which are rotationally invariant that is, $E$ and $B$. As mentioned before, $E$ 
and $B$ behave differently under parity transformation; $E$ remains 
unchanged whereas $B$ changes sign.
This is analogous to the directionality of electric and magnetic fields,
which are proper and pseudo vectors, respectively.

Following \cite{Bran_eb_1}, we define $R= E +iB$.
We discuss the details of the $E$ and $B$ decomposition from linear 
polarisation in the small-scale limit and the two sign conventions in
Appendix~\ref{app1}.
The sign convention adopted here agrees with that of \cite{Bran_eb_1},
but is different from the one in \cite{Bran_eb_2}, who followed the convention of \cite{Durrer}.
In the small-scale limit, $R$ is related to $P$ in Fourier space
(indicated by tildes) via the following relation
\citep[for details, see][]{Zaldarriaga,Seljak_small}
\begin{equation}
	\tilde{R}(k_x,k_y) = (\hat{k}_x - i\hat{k}_y)^2 \tilde{P}(k_x,k_y),
	\label{small_scale_rel}
\end{equation}
where $\hat{k}_x$ and $\hat{k}_y$ are $x$ and $y$ components (in the plane 
of the image) of the unit vector $\hat{\vec{k}} = \vec{k}/k$,
where $k=|\vec{k}|$ is the length of $\vec{k}=(k_x,k_y)$.
Upon transformation of $\tilde{R}$ back to real space, we have maps of $E(x,y)$ 
and $B(x,y)$ corresponding to a set of $Q$ and $U$ maps at a certain 
wavelength. It is useful to compute shell-integrated spectra in wavenumber 
space for a given radius $k$ as
\begin{equation}
	C_{XY}^i(k) = \int_0^{2\pi} \tilde{X}_i(\vec{k})\tilde{Y}_i^\ast(\vec{k}) \hspace{0.1 cm}k \hspace{0.1 cm} \dd \phi_k,
  \label{shell_int}
\end{equation}
where the asterisk denotes complex conjugation,
$\tilde{X}_i$ and $\tilde{Y}_i$ stand for $\tilde{E}_i$, $\tilde{B}_i$, or $\tilde{T}$ ($T$
represents temperature),
and $i$ characterises the wavelength bin at which we compute $E$ and $B$.
For $T$ we take the continuum intensity as a proxy, so there is no subscript $i$.
Finally, we also define the normalised antisymmetric spectra as
\begin{equation}
  c_{\rm A}^{XY}(k) = \frac{\sum_{i} 2C_{XY}^i(k)}{\sum_{i}[C_{XX}^i(k)+C_{YY}^i(k)]},
\label{antisym}
\end{equation}
which we use throughout our analysis; see Appendix~\ref{app2} for examples.

We considered between four and six wavelength bins, as is
explained in more detail in Sect.~\ref{data1}.
However, in some cases (Sect.~\ref{magEB}), we inferred Stokes $Q$ and $U$
from the components of the transverse magnetic field, in which case
there is no wavelength dependence.

\section{Application to solar observations} \label{data}

\subsection{Observations used in this study}\label{data1}

In this section, we briefly describe the solar observations 
from the Helioseismic and Magnetic Imager \cite[HMI,][]{Schou}, on board 
{\em SDO}, used in this analysis. We used the publicly available 
polarisation measurements and 
magnetic field vector data at different stages in our analysis. For the
polarisation measurements, we used level-1 reduced data. Here we
only focus on Stokes $Q$ and $U$; Stokes $I$ and $V$ are not included in our
study. {\em SDO}/HMI provides full disk images of Stokes $Q$ and $U$ 
which were cropped to the relevant ARs. The magnetic field
vector data are the result of the VFISV inversion code \citep{Borrero}, 
and a disambiguation based on the minimum energy method \citep{Metcalf94,Leka09}.
We used the SHARP data product \citep{Bobra} from
the HMI team, which provides the definitive $\vec{b}=(b_r,b_\theta,b_\phi)$ which has been
remapped to a Lambert Cylindrical Equal-Area projection and decomposed
into $b_r$, $b_\theta$, and $b_\phi$. These however are not full disk,
but partial-disk patches, automatically identified and cropped around
the ARs.

\begin{table}
\caption{List of ARs used in this study. The last two
columns are the complexity of ARs and category the ARs fall into
based on our analysis.}
\label{table:1}      
\centering                          
\resizebox{\columnwidth}{!}{
\begin{tabular}{c r c r c c} 
\hline\hline                
NOAA no. & Date$\quad$ & Hemis. & Lat.\ [$\degr$]$\!\!\!$ & Comp. & Cat.\\   
\hline                        
   12042 & 21/04/2014 & North & $18.4$ & $\beta$ & A \\
   12158 & 11/09/2014 & North & $14.1$ & $\beta\gamma$ & A \\     
   12090 & 16/06/2014 & North & $24.0$ & $\beta$ & A\\
   11546 & 22/08/2012 & North & $15.5$ & $\alpha$ & A \\
   11117 & 25/10/2010 & North &  $1.1$ & $\beta$ & A\\
   11486 & 24/05/2012 & North & $15.0$ & $\beta$& A\\
   11543 & 13/08/2012 & North & $21.3$ & $\beta\gamma$ & B\\
   12022 & 2/04/2014 &  North & $17.3$ & $\alpha$ & C\\
   12387 & 20/07/2015 & North & $13.7$ & $\beta$ & C \\
   \hline
   12186 & 13/10/2014 & South & $-20.5$ & $\alpha$ & A \\
   11542 & 12/08/2012 & South & $-13.5$ & $\beta$ & A \\
   12418 & 18/09/2015 & South & $-17.3$ & $\beta$ & A \\
   11490 & 29/05/2012 & South & $-12.5$ & $\beta$& A\\
   12045 & 25/04/2014 & South & $-24.0$ & $\beta$ & A\\
   12075 & 29/05/2014 & South &  $-9.0$ & $\alpha$ & A\\
   12415 & 16/09/2015 & South & $-21.1$ &$\beta\gamma$ & B \\
   12194 & 26/10/2014 & South & $-12.0$ & $\alpha$ & B \\
   11494 &  6/06/2012 & South & $-19.7$ & $\beta$ & C \\
\hline                                  
\end{tabular}}
\end{table}
We chose a small, random sample of
ARs from solar cycle $24$ (see Table~\ref{table:1}).
We examined the antisymmetric polarisation correlations (see Sect.~\ref{theory}),
$c_{\rm A}$, calculated from Stokes $Q$ and $U$ measurements of the ARs.
To reiterate, the aim is to check whether we see a systematic 
preference for the sign of $c_{\rm A}$ based on hemisphere, thus reflecting 
the hemispheric sign rule for magnetic helicity.

HMI is a filtergraph which samples the $6173\Ang$ \ion{Fe}{i} absorption line
at 6 positions in wavelength with a spacing of $69\mA$. 
The full width half maximum of the filter at each of these wavelengths is
$76$~m\AA{}~$\pm 10$~m\AA{}, we therefore refer to them as
wavelength bins $\lambda_i$, where
$i = 0$ to $5$. Here $\lambda_0$ is the extreme blue position of the
filter, and $\lambda_5$ is the extreme red position.

We produced maps of Stokes $Q$ and $U$ at all wavelength bins within
the $6173\Ang$ \ion{Fe}{i} line, on both the blue and red wings.
From the Stokes maps we then computed the $E$ and $B$ polarisations using
Eq.~(\ref{small_scale_rel}) and also the shell-integrated spectra, Eq.~(\ref{shell_int}),
at these wavelength bins. As mentioned before, Faraday rotation can possibly
contribute to a non-zero $EB$ correlation, even in the absence of magnetic helicity.
Its effects are strongest near or at the line core, depending on 
the strength of the magnetic field. We studied these ARs on the central meridian,
so the Doppler shifts due to solar rotation 
and Evershed flow
are minimised. However, there are also Doppler
shifts due to the orbital velocity of {\em SDO}, 
resulting in the line core  
being sampled by $\lambda_2$ or $\lambda_3$. We note that
in extreme cases these shifts due to the orbital velocity could be large
enough for the line core to be sampled by $\lambda_1$ or $\lambda_4$.
For these reasons, we
obtain $c_{\rm A}$ from the averaged $E$ and $B$ spectra computed at $\lambda_0$,
$\lambda_1$, $\lambda_4$, $\lambda_5$ and 
analyse the $EB$ correlation from $\lambda_2$ and $\lambda_3$
separately (Sect.~\ref{lam23}). 

We separated the 18 ARs of Table~\ref{table:1} into three categories based on the
sign of the $EB$ correlation $c_{\rm A}^{EB} (k)$.
Category~A (12 ARs) is for ARs whose normalised $c_{\rm A}^{EB} (k)$ spectra
show a preference for a particular sign that is in agreement with the expected hemispheric
sign rule for magnetic helicity (see Sect.~\ref{intro}).
Category~B (3 ARs) is for ARs that show the opposite sign for $c_{\rm A}^{EB} (k)$ than 
what is expected from theoretical considerations. Finally, category~C (3 ARs) is for ARs that
do not show any clear preference for the sign of $c_{\rm A}^{EB} (k)$. 
The dataset containing the shell-integrated spectra defined in Eq.~(\ref{shell_int}), along with maps of $E$ and $B$ for each AR can be found in an online catalogue\footnote{\url{https://doi.org/10.5281/zenodo.3888575}}.

\subsection{ARs from category A}\label{data2}

\begin{figure*}
    \centering
    \includegraphics{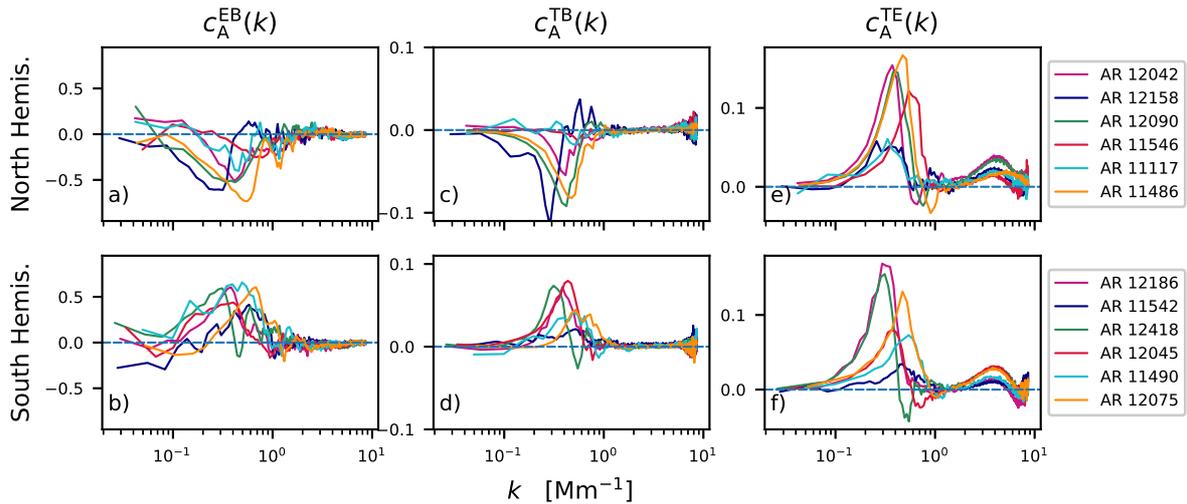}
    \caption{$c_{\rm A}^{EB} (k)$, $c_{\rm A}^{TB} (k)$, and $c_{\rm A}^{TE} (k)$
        (first, second, and third column) for the ARs of category~A from the Northern  
 	hemisphere (top row) and Southern hemisphere (bottom row) (see 
 	Table~\ref{table:1}), using Eq.~(\ref{antisym}) with $E$ and $B$
        being computed at $\lambda_0,\lambda_1,\lambda_4,\lambda_5$.}
    \label{avg_maj_NH}
\end{figure*}

Firstly, we present in Fig.~\ref{avg_maj_NH} (first column) the spectra $c_{\rm A}^{EB} (k)$ of
correlations of $E$ and $B$ calculated from Stokes $Q$ and $U$
at four wavelengths ($\lambda_0,\lambda_1,\lambda_4,\lambda_5$); see Eq.~(\ref{antisym}).
In our analysis, we find that the ARs from the Northern (Southern) 
hemisphere, have a preference for a negative (positive) sign of
$c_{\rm A}^{EB}$. The non-zero $c_{\rm A}^{EB}$
correlations computed from those wavelength bins where
the influence of Faraday rotation should be negligible, suggest that these
correlations are indeed a proxy for helical magnetic fields.

As discussed in Sect.~\ref{intro}, helical 
magnetic fields can contribute to parity-odd correlations.
In addition to correlations between $E$ and $B$, there can also
be parity-odd correlation from $T$ and $B$ ($c_{\rm A}^{TB}$),
temperature being a parity-even quantity. 
We used the continuum intensity, which is an excellent proxy for the temperature of 
the photosphere. Figure~\ref{avg_maj_NH} (middle column) shows the
resulting spectra, which are an average of four wavelength bins
($\lambda_0,\lambda_1,\lambda_4,\lambda_5$). 
In accordance with $c_{\rm A}^{EB}$, one observes a 
hemispheric sign preference for $c_{\rm A}^{TB}$ correlations: negative 
(positive) in the Northern (Southern) hemisphere. The non-zero values of 
$c_{\rm A}^{TB}$  along with the systematic preference for the sign is yet 
another indicator that these antisymmetric correlations 
are a result of helical fields. 
 This sign preference is especially prominent for scales between 1 and 10~Mm.

By studying the analogously computed correlations of $T$ and $E$
($c_{\rm A}^{TE}$), we can have another confirmation that it is indeed
$B$ that is changing sign with hemisphere and not $E$.
The $c_{\rm A}^{TE}$ correlations (Fig.~\ref{avg_maj_NH}, third column)    
mostly maintain the same positive sign for ARs from both hemispheres, which
is expected given the parity-even nature of $E$ (see Appendix~\ref{app1}).

\subsection{ARs from category B}\label{data3}

Out of the ARs that we looked at in this
study, we also found ARs that show an opposite sign 
of $c_{\rm A}^{EB}$ than what is expected from the hemispheric
sign rule (Fig.~\ref{avg_min_NH}, solid lines). 
We expect the sign for an AR in the North
(South) to have a negative (positive) sign of magnetic helicity.
ARs~11543 (North) and 12415, 12194 (South) show opposite signs, 
positive and negative, respectively. This is not surprising as such, since 
in most statistical studies that look at helicity of
isolated patches of ARs, there is always a certain percentage of
ARs that do not conform to the hemispheric sign rule for
helicity \citep{Pevtsov1995,Singh18,Gosain}. The latitude 
or the complexity class of the ARs does not seem to
play a role in the reversed sign the $EB$ correlations,
as ARs of a similar latitude or complexity also belong to category~A.

For the category~B cases, analogously to the category~A cases, 
we also looked at correlations between $T$ and $B$, $c_{\rm A}^{TB} (k)$, 
and also between $T$ and $E$, $c_{\rm A}^{TE} (k)$.
First, looking at $c_{\rm A}^{TB} (k)$ (Fig.~\ref{avg_min_NH},
second column), AR~11543 displays a distinctly bimodal 
behaviour in that there are positive and negative values of
$c_{\rm A}^{TB} (k)$ at slightly different values of $k$.
This is in contrast to ARs of category~A, where the $TB$ correlations
showed a clear preference for a particular sign, in accordance with 
the sign of $EB$ correlations. By contrast, for ARs~12415 and 12194, 
the $EB$ and $TB$ correlations have the same negative sign.
As previously, the $TE$ correlations (Fig.~\ref{avg_min_NH}, third column) 
are positive for both hemispheres and in the case of AR~11543, $c_{\rm A}^{TE} (k)$ 
shows an unusual double-peaked spectrum.

\subsection{ARs from category C}\label{data4}

\begin{figure*}
    \centering
    \includegraphics{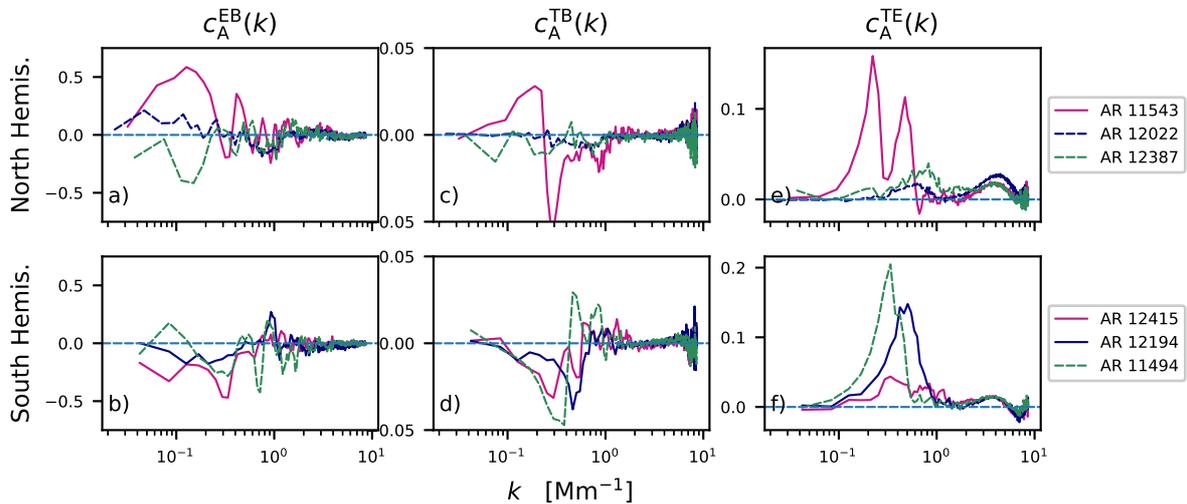} 
    \caption{Similar to Fig.~\ref{avg_maj_NH}, but the ARs of
    categories~B (solid lines) and C (dashed lines).}
    \label{avg_min_NH}
\end{figure*}

The third category is for ARs that do not show a clear preference
for a sign (Fig.~\ref{avg_min_NH}, dashed lines) of $c_{\rm A}^{EB}$.
Looking at $TB$ correlations for this category, 
the two
ARs from the Northern hemisphere
display almost no signal. However, the $TB$ correlations for 
the Southern
AR~11494 show a bimodal behaviour,
similar to AR~11543 from category~B.
A similar hemispheric distinction can also be seen for the $TE$ correlations, 
where for the Northern ARs there is almost no signal, 
and for the Southern AR we have a clear positive sign.

\subsection{$EB$ correlations computed from the magnetic field}\label{magEB}

Until now, we have decomposed the measured Stokes $Q$ and $U$ into $E$ and $B$
polarisations. Here, following \cite{Bran_eb_2}, we made an attempt  
to compute $E$ and $B$ (thus also $c_{\rm A}^{EB}$) from the magnetic field 
data. 
This is because the components of the magnetic field vector, used to compute $c_{\rm A}$,
are from a spectropolarimetric inversion wherein the magneto-optic
effects and Doppler shifts are accounted for.
If we observe a region closer to the center of the solar disk, we can, to a 
certain degree, assume that
\begin{equation}
P \equiv Q+iU = - \epsilon(b_\theta+ib_\phi)^2,
\label{PfromB}
\end{equation}
where $\epsilon$ in the present context is a proportionality constant that depends on the heliocentric angle
and $b_\theta, b_\phi$ are the transverse field components in the medium.
Equation~(\ref{PfromB}) is an approximation of the otherwise complex relation 
between Stokes $Q$ and $U$ to the transverse components $b_\theta, b_\phi$.
There are two things that we must note here. Firstly, we 
are assuming that the Stokes $Q$ and $U$ signals are only due to the 
magnetic field components parallel to the solar surface (transverse components). 
However, this is valid only at low heliocentric angles; farther away from
the disk center the validity of this assumption is poor.
Secondly, as pointed out in 
\cite{Bran_eb_2}, the $\pi$ ambiguity associated with the 
transverse components $(b_\theta ,b_\phi)$ does not affect this assumption, 
that is, a flip of $180^{\circ}$
of the transverse component does not change the sign of $P$.
When we compute the $c_{\rm A}^{EB}$
from the magnetic field, we do this from maps of
$b_\theta$ and $b_\phi$ by exploiting Eq.~(\ref{PfromB}).
In the following, since we are only interested in normalised quantities such
as $c_{\rm A} (k)$, which are relative measurements, we put $\epsilon=1$.

\begin{figure*}
    \centering
    \includegraphics{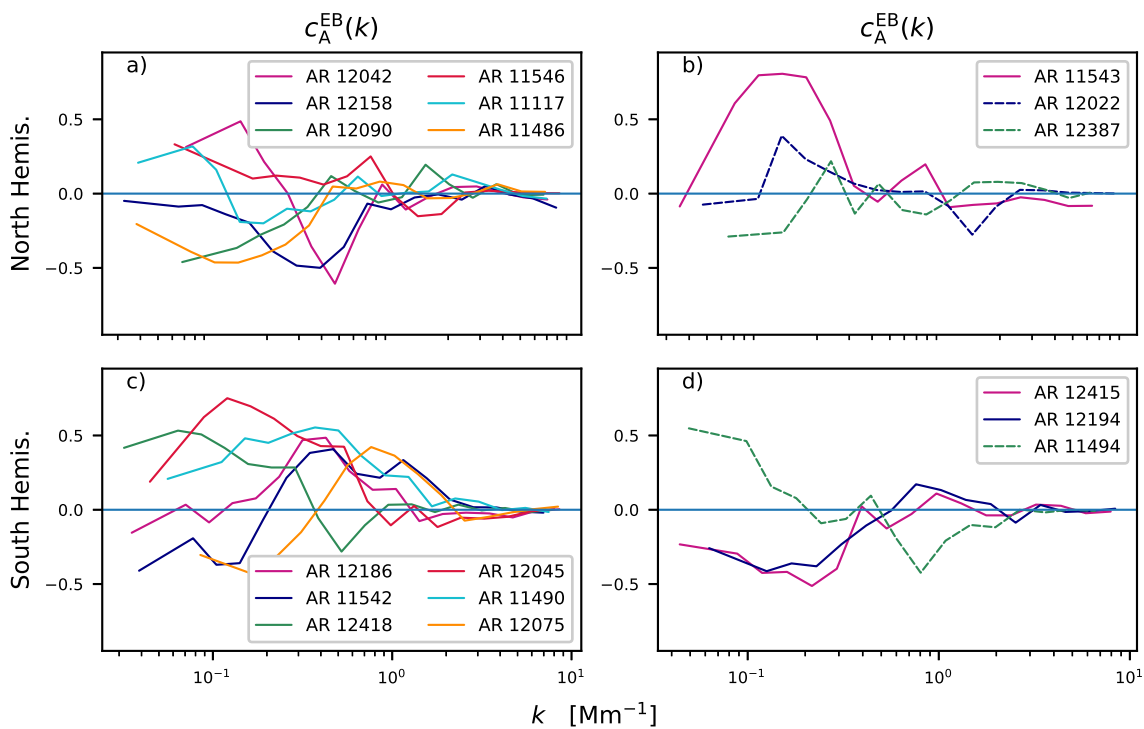}
    \caption{$c_{\rm A}^{EB} (k)$ for ARs of categories~A (left column),
    B (right column, solid lines), and C (right column, dashed lines), with $E$
    and $B$ being calculated from the components of the magnetic
    field vector. The curves for $c_{\rm A}^{EB} (k)$ are 
    smoothed in logarithmically spaced bins, for 
    better visibility.}
    \label{B_maj}
\end{figure*}

We show in Fig.~\ref{B_maj} the spectrum $c_{\rm A}^{EB}$
computed from $b_\theta$ and $b_\phi$ for all ARs.
First we look at ARs of category~A (Fig.~\ref{B_maj}, left column).
The preference for a negative (positive) sign of $c_{\rm A}$ in the
Northern (Southern) hemispheres is evident, although it is definitely
less clear than when $c_{\rm A}^{EB}$ is computed directly from linear
polarisation (Fig.~\ref{avg_maj_NH}, first column). 
This is especially true of the case of AR 11546, which would be classified as an AR of category~B,
if one looks at $c_{\rm A}^{EB}$ computed from the magnetic field (see left panel of Fig.~\ref{B_maj})
with the simplifying assumption mentioned in the paragraph above.
This weaker preference can be attributed to the imperfect validity of Eq.~(\ref{PfromB}) at AR
latitudes
further away from the equator, since the linear polarisation in this case has  
a significant additional contributions from $b_r$.
For category~B (Fig.~\ref{B_maj}, right column, solid lines), the preference
for the reversed sign of $EB$ correlations is also quite discernible.
And lastly, for category~C, even for the $EB$ correlations computed using Eq.~(\ref{PfromB}),
an obvious preference for either of the signs is absent.
However, regardless of the categories, one can notice good agreement in the shape of the spectra
of individual ARs computed from the magnetic field
(Fig.~\ref{B_maj}) and those computed from Stokes $Q$ and $U$
(Figs.~\ref{avg_maj_NH} and \ref{avg_min_NH}, first column).
This agreement between the spectra is an indication that the
$c_{\rm A}$ correlations we see from Stokes $Q$ and $U$ are indeed
indicative of the intrinsic magnetic helicity of the
ARs, and not a result of Faraday rotation from a non-helical
magnetic field.

\subsection{$EB$ correlations near line core}\label{lam23}

In the previous section, we mainly looked at the various correlations
($EB$, $TB$, $TE$) computed at wavelength bins
$\lambda_0, \lambda_1, \lambda_4,$ and $\lambda_5$.
To minimise the influence of Faraday rotation on these correlations,
we intentionally left out $\lambda_2$ and $\lambda_3$,
which are at or closest to the line core.
We recall that Faraday rotation can, in principle,
contribute to parity-odd correlations, even in the absence of 
helical magnetic fields. In this section,
we take a closer look at the $EB$ (and other) correlations
from $\lambda_2$ and $\lambda_3$. 

\begin{figure*}
    \centering
    \includegraphics{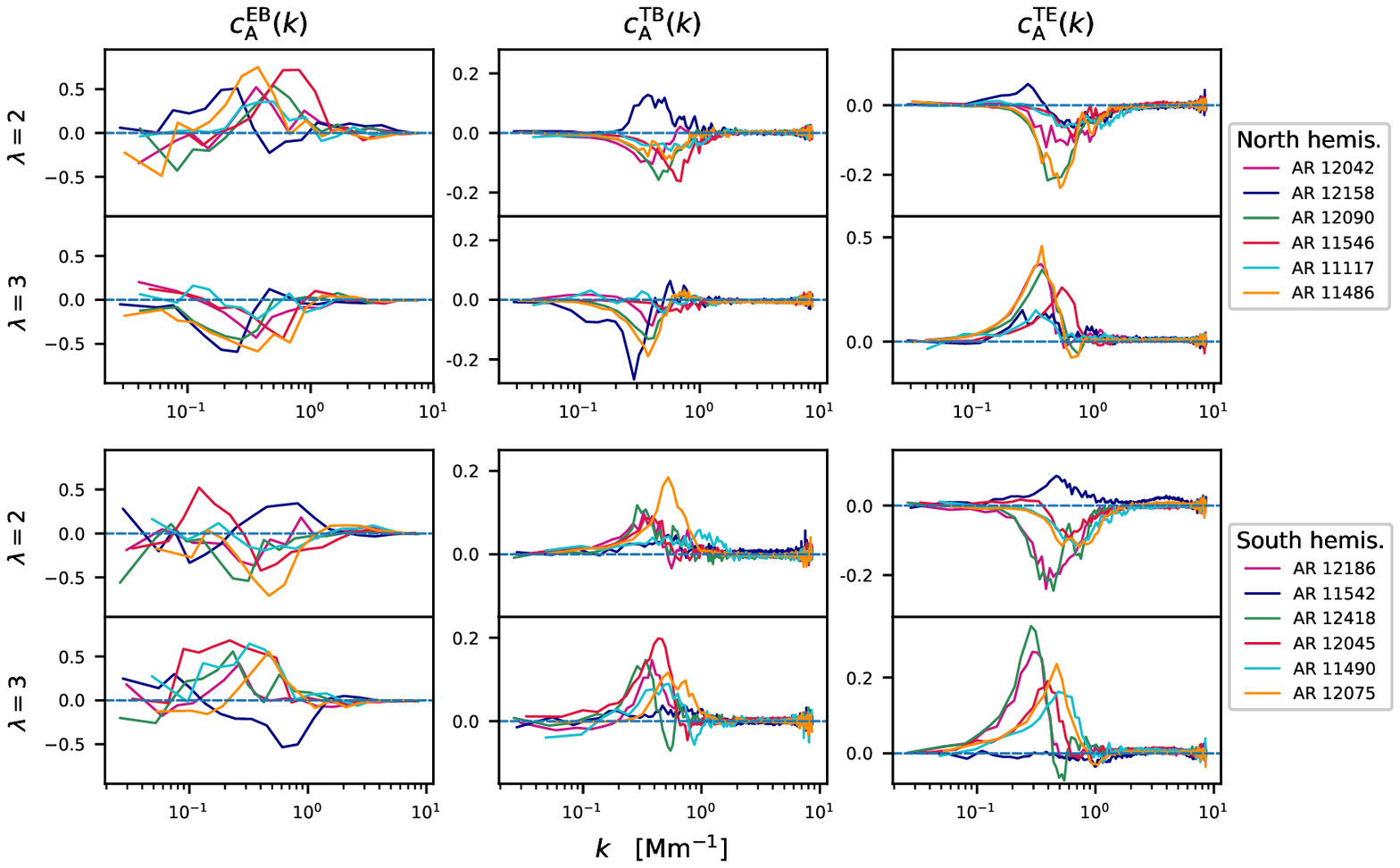}
    \caption{$c_{\rm A}^{EB} (k)$, $c_{\rm A}^{TB} (k)$, and $c_{\rm A}^{TE} (k)$
  (first, second, and third column) for the ARs of category~A (see 
  Table~\ref{table:1}) from the Northern  
  hemisphere (top row) and Southern hemisphere (bottom row),
  using Eq.~(\ref{antisym}) with $E$ and $B$ being computed
  at $\lambda_2,\lambda_3$.}
    \label{lam_2_3_maj_NH}
\end{figure*}
 
\begin{figure*}
   \centering
   \includegraphics{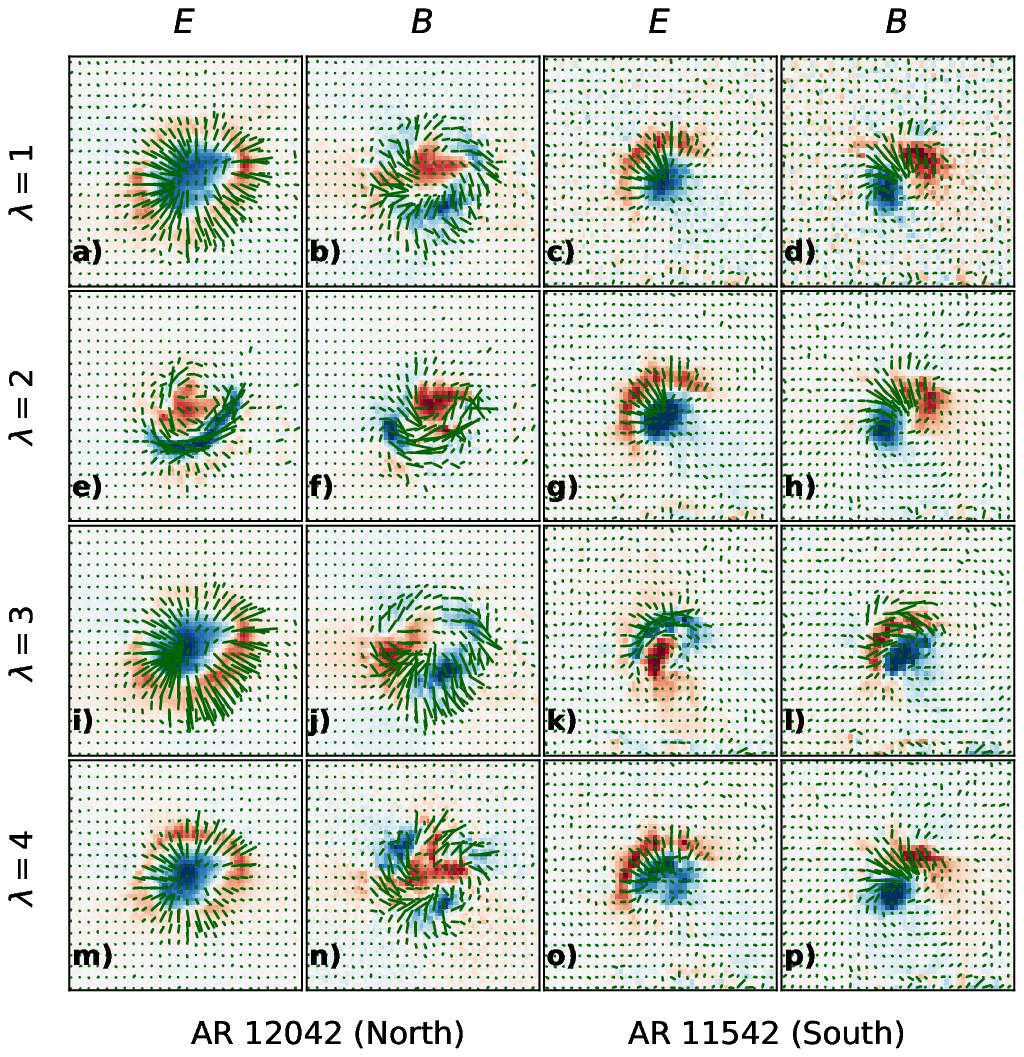}
   \caption{$E$ and $B$ maps for two ARs, computed
   from Stokes $Q$ and $U$. The pattern of polarisation (green lines) have been scaled
   according to total linear polarisation signal ($\sqrt{Q^2+U^2}$). The colour scale is
   has been adjusted for each frame to easily see the positive (red) and negative (blue) values 
   of $E$ and $B$.}
   \label{EB_maps}
\end{figure*}

From our analysis in the previous section, we find that for ARs of category~A, 
$c_{\rm A}^{EB}$ and $c_{\rm A}^{TB}$ have 
a negative (positive) sign in the Northern (Southern) hemisphere
which conforms to the expected hemispheric sign rule for magnetic helicity.
We first take a look at the correlations computed from the
wavelength bin $\lambda_2$. From Fig.~\ref{lam_2_3_maj_NH}, we see that,
at $\lambda_2$, $c_{\rm A}^{EB}$ shows a sign reversal in both hemispheres, 
positive (negative) in the North (South). Surprisingly enough, $c_{\rm A}^{TB}$
does not show this sign reversal, its signs in both hemispheres are consistent
with our analysis (except for the peculiar case of AR~12158). 
This curious behaviour is better understood when one looks
at the $TE$ correlations from this wavelength bin. Based on our analysis we
know that $c_{\rm A}^{TE}$ shows a positive sign (see Fig.~\ref{avg_maj_NH},
 third column) in both hemispheres ($E$ is parity-even).
At $\lambda_2$, however, $c_{\rm A}^{TE}$ is negative in both
hemispheres. Thus, the sign reversal in $c_{\rm A}^{EB}$ at $\lambda_2$ is 
a result of $E$ changing sign. We investigated whether Faraday rotation is causing
this sign change in the next subsection.

The inspection of the $\lambda_3$ bin (Fig.~\ref{lam_2_3_maj_NH} second and fourth row) reveals that
the peculiar sign reversal of $EB$ correlations
is absent for most ARs of category~A. The sign of $c_{\rm A}^{EB}$ is 
consistent with our previous analysis, except for AR 11542,
for which the sign is negative and opposite to that expected for an AR in the South.
The signs of $c_{\rm A}^{TB}$ in this bin are also consistently negative 
(positive) in the North (South), as seen before. The same is true for 
$c_{\rm A}^{TE}$, which is positive in both hemispheres for most ARs.
The sign reversal in the $EB$ correlations for AR 11542 is
due to a change in sign of $E$ rather than $B$, if we take a close look
at its $TE$ correlations. 
For categories~B and C, we see the same reversal of $E$ to negative signs
at $\lambda_2$ (not shown), except for ARs~12022 and 12387, where the amplitudes
of the correlations are too low to discern a sign reversal. This indicates that,
regardless of the categories of the ARs, $E$ changes sign in the wavelength
bins closest to or at the line core.

In Fig.~\ref{EB_maps}, we show maps of $E$ and $B$
for two ARs.
We have just seen that at $\lambda_2$, $c_{\rm A}^{EB}$ shows a 
reversed sign to positive (instead of negative) for an AR in the North, and
we can infer that this sign reversal is due to a change in the sign of $E$. 
Figure~\ref{EB_maps}e also shows a different (positive in this
case) sign of $E$ in the center of the AR compared to the
other wavelength bins; cf.\ Figs.~\ref{EB_maps}a,i,m. A similar behaviour
can be seen for AR~11542 at $\lambda_3$; see Fig.~\ref{EB_maps}k.
There is a change in sign of $E$ (positive again), which corresponds to a change
in $c_{\rm A}^{EB}$ at $\lambda_3$; see Fig.~\ref{lam_2_3_maj_NH}.
We find that from our sample of ARs, almost all ARs display
a sign reversal of $EB$ and $TE$ at $\lambda_2$ and for one
AR 11542 at $\lambda_3$.
In the case of AR 11542, we found the 
spectral line to be red-shifted as compared to the other observations,
which explains the sign reversal at $\lambda_3$ instead of $\lambda_2$.
Thus, the mechanism 
causing the change in sign of $E$,
can affect both wavelength bins, $\lambda_2$ or $\lambda_3$, depending on the
Doppler shift of the spectral line.

\subsection{Tests for effects of Faraday rotation}\label{test}
\label{FaradayRot}

Faraday rotation can change the different states of linear polarisation
amongst themselves. Therefore, at certain wavelengths within
a spectral line, depending on the magnetic field strength,
the effects of Faraday rotation are the strongest. At these
wavelengths, the maps of Stokes $Q$ and $U$ show a swirl-like pattern
because of the different states of linear polarisation getting
interchanged amongst themselves. $B$ polarisation is
sensitive to a curl-type pattern; therefore even a non-helical field,
due to Faraday rotation, can give rise to
a non-zero $c_{\rm A}^{EB} (k)$.
This effect of Faraday rotation has been examined for dynamo-generated helical magnetic
fields in a sphere \citep{Bran_eb_2} using Eq.~(\ref{PfromB}).

\begin{figure}
    \centering
    \includegraphics{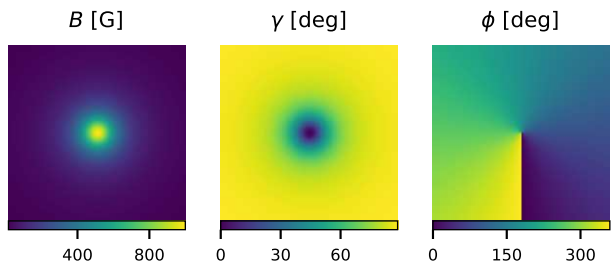}
    \caption{Magnetic field strength, inclination($\gamma$) and azimuth ($\phi$) for a 
    simplistic sunspot-like configuration at the disk center.}
    \label{b_config}
\end{figure}

\begin{figure*}
    \centering
    \includegraphics{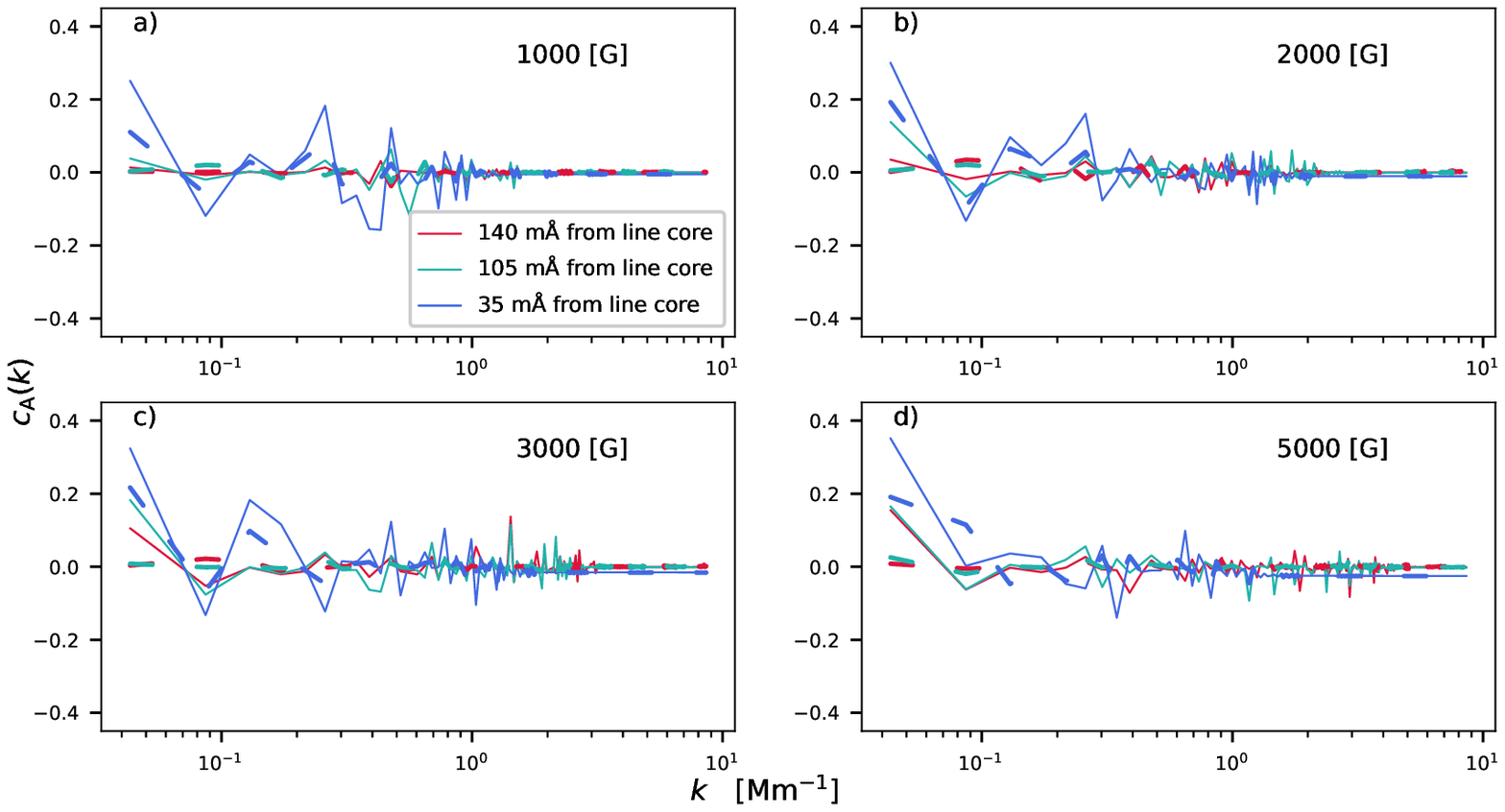}
    \caption{$c_{\rm A}^{EB} (k)$ calculated from Stokes $Q$ and $U$ for the 
    synthetic test cases. The four panels correspond to different field 
    strengths. The solid lines are for the spot-like configuration at the 
    disk center. The dot-dashed lines are for the spot configuration at $30^
    \circ$ latitude on the solar disk. 
    For all the synthetic
    cases, we chose to assign the plate scale of HMI -- hence the abscissa is
    in ${\rm Mm}^{-1}$.}
    \label{syn_fararot}
\end{figure*}

In this section, we describe some simple tests we carried out to
isolate the contribution of Faraday rotation from a non-helical
magnetic field to the $c_{\rm A}^{EB} (k)$ correlation, when one computes
it from Stokes $Q$ and $U$ near or at the line core.
We started with a simple model of the
solar atmosphere, 
the temperature
stratification is based on the Harvard Smithsonian Reference Atmosphere \citep{Gingerich71}.
We introduced a magnetic field configuration which is constant with height
in our model atmosphere; see Fig.~\ref{b_config}.
The field strength is decreasing outwards
from the center following a Lorentzian profile, the inclination
($\gamma$) with respect to the line-of-sight was chosen in a way to
make the magnetic field diverge away from the center, and
similarly the azimuth was chosen such that the field was uniformly
distributed in the transverse plane. We used {\tt STOPPRO} \citep{Solanki1987},
a numerical code which solves the RTE to synthesise the full
Stokes vector for the $6173 \Ang$ \ion{Fe}{i} absorption,
with a spectral resolution of $5\mA$. 
The spectra were
synthesised for two distinct cases: one when such a magnetic
configuration is at the disk center and another where it is at
$30^\circ$ in latitude. This latitude is roughly coinciding with the ones of the ARs
we looked at in our analysis.
The synthetic spectra were not degraded to account for instrumental effects and
since we do not have any velocity gradients in our model,
the spectra are symmetric about the line core. 
In both cases, we investigated
different field strengths, keeping the inclination and azimuth of the
magnetic field vector the same. We chose such a distribution of the
magnetic field to mimic the magnetic field of a sunspot and 
to make sure that this field configuration is non-helical. This way, any
non-zero correlation can only
be attributed to Faraday rotation. 
Figure~\ref{syn_fararot} shows the
$c_{\rm A}^{EB} (k)$ at three different wavelengths, 
the one closest to the line core roughly falls in the $\lambda_2$ bin
and the wavelength $105\mA$ away in the $\lambda_1$ bin.
Since the spectra are symmetric, the $c_{\rm A}^{EB} (k)$ at $\lambda_3,\lambda_4$
is identical to the ones shown in the figure.
We find the highest amplitudes of $c_{\rm A}^{EB}$ when the
magnetic field configuration (regardless of the field strength) is at
the disk center (solid lines in Fig.~\ref{syn_fararot}) as compared to
when it is viewed from a position at $30^\circ$ in latitude
(dash-dotted lines in Fig.~\ref{syn_fararot}). 
In all cases, the maximum amplitude of $|c_{\rm A}^{EB}|$ for $k$ between $0.1$ to $1$
is not higher than $0.2$, but when it is computed from observations it is
around $0.5$ (see Fig.~\ref{avg_maj_NH}). Also, unlike observations, $c_{\rm A}^{EB}$
fluctuates around zero for these test cases.
In terms of wavelength, the largest contributions from Faraday rotation to 
$c_{\rm A}^{EB}$
are, as expected, from wavelengths closest to the nominal line core.

Now we turn our attention to the sign reversal of $E$ (and consequently of $EB$ and $TE$).
The $E$ maps in Fig.~\ref{EB_maps} show a sign reversal at $\lambda_2$ for AR~12042 
(panel e) and $\lambda_3$ for AR~11542 (panel k). It is tempting to relate this
sign reversal to the effect of Faraday rotation, which is strongest at or close to
the line core. Depending on the orbital velocity of {\em SDO}, this maximum can fall into
wavelength bins $\lambda_2$ or $\lambda_3$. However, the $E$ maps computed from
the simple model of the solar atmosphere described above do not show any hint of 
a sign reversal. This means that either our model is too simple and not representative of the
observations presented in Fig.~\ref{EB_maps}, or that Faraday rotation is not the 
mechanism responsible for the sign reversal. 
We performed three experiments to examine this further.

In the first experiment we increased the complexity of the model by adding a filamentary 
fine structure directed radially outwards from the center of the synthetic spot, 
representative of a penumbra.
But also this model failed to reproduce the sign reversal of $E$.
In a second experiment we investigated the effect of a vertical gradient in the magnetic field
parameters, which  were neglected in the simple model described above. We 
computed the response functions of the $Q$ and $U$ profiles with respect to 
variations of the magnetic field strength in a typical umbral and penumbral 
atmosphere. The response functions describe the wavelength and height 
dependence of the Stokes parameters.  We found the $Q$ and $U$ profiles to be 
sensitive over a $\approx$200\,km thick layer above the optical depth unity 
surface with a rather uniform wavelength dependence. 
With typical gradients in a sunspot of $\approx$1\,G/km and the absence
of a significant wavelength dependence, this height difference is 
too small to produce a large enough Faraday rotation and, therefore, it also does
not explain the observed sign reversal in the $E$ maps.


The third experiment is based on the data underlying  Fig.~\ref{EB_maps}. 
We applied the Milne-Eddington inversion code HeLIx+ \cite[]{Lagg04,Lagg09} 
to the HMI data of AR~12042 to retrieve its atmospheric parameters 
(e.g. magnetic field vector, line-of-sight velocity). The inversions 
reproduce the observed profiles reasonably well, the $E$ and $B$ maps 
computed from these fitted profiles are very similar to the observed 
$E$ and $B$ maps and show the observed sign change mostly at $\lambda_2$. 
In a next step, we used the atmospheric parameters from this inversion
to compute synthetic Stokes profiles, neglecting the effect of Faraday 
rotation. The $E$ and $B$ maps computed from these profiles are very 
similar to the maps including Faraday rotation and show the sign change equally
well. This proves that the observed sign change cannot be a result of
Faraday rotation.

\section{Conclusions}\label{dis}
The study is motivated by an earlier work by \cite{Bran_eb_1}, who
demonstrated that the $EB$ decomposition of linear polarisation
can, under inhomogeneous conditions, be a proxy for magnetic helicity. 
However, they did not retrieve any significant non-zero $EB$ correlations 
when they tested this proxy with solar observations from VSM/SOLIS. 
In this work, we looked at individual ARs from both hemispheres, 
observed with {\em SDO}/HMI, and not only recovered significant 
$EB$ correlations, but also a systematic dependence 
of its sign on hemisphere. We found the sign of both $EB$ and $TB$
correlations to be consistent with that of small-scale magnetic helicity, 
that is, negative (positive) in the Northern (Southern) hemisphere.
We note that this is opposite to what was reported in \cite{Bran_eb_1},
based on numerical simulations of rotating convection.
We also found that such parity-odd correlations ($EB$, $TB$), 
which are a good proxy for magnetic helicity, can be reliably 
computed from linear polarisation away from the line core of spectral lines.
This minimises the influence of Faraday rotation on the 
various correlations and, since we use polarisation measurements directly,
circumvents the $\pi$ ambiguity.
We found this to be true for the
majority of the ARs we looked at (12 out of 18, category~A).
We also found 3 ARs (category~B) that showed a reversed sign for the
parity-odd correlations compared to what is predicted by theory, and 3 ARs
(category~C) that displayed no preference for a specific sign.
As already alluded to above, the existence of ARs that display a reversed
sign of parity-odd correlations (category~B) may not be surprising since 
it has been demonstrated in previous studies \citep{Pevtsov1995,Singh18,Gosain} 
that there is a certain percentage of ARs that are 
in violation of the hemispheric sign rule for magnetic helicity. 
Proximity to the solar equator, complexity of the ARs,
among others, are speculated to be the possible reasons behind 
these violations. Based on our present studies, the position of the AR
on the solar disk or its complexity
do not seem to be a factor in classifying an AR into 
categories~A or B (see Table~\ref{table:1}).
However, our sample size is too small to draw more robust conclusions 
about this, so a more systematic study with a larger sample size is desirable.

We also computed the $EB$ correlations from the transverse ($b_\theta$, $b_\phi$)
components of the magnetic field vector (Sect.~\ref{magEB}). The correlation
$c_{\rm A}^{EB} (k)$ retrieved from this approximation matched the shape
of the spectra retrieved directly from linear polarisation. A
preference for a particular sign was less clear when $c_{\rm A}^{EB}$
was computed from the magnetic field, because the validity of
Eq.~(\ref{PfromB}) is questionable
at AR latitudes. Nevertheless, it still gives us another
confirmation that non-zero amplitudes of $c_{\rm A}^{EB}$ are not
due to Faraday rotation, since the magnetic field is inferred after
accounting for magneto-optical effects. \cite{Bran_eb_2} also 
used the transverse components of the
magnetic vector to compute $EB$ correlations on a global scale.
By using spin-2 spherical harmonics to compute $E$ and $B$ polarisations,
and a heuristic approach to account for North-South sign change of
magnetic helicity, he could successfully retrieve maximum power
at the smallest wavenumbers. This is possibly due to the fact that
the $EB$ decomposition approach (regardless of it being computed from
magnetic field or polarisation) is insensitive to the disambiguation,
which can affect correlations at large scales, where field strengths are weak.

In Sect.~\ref{lam23}, we looked at the different correlations 
computed from linear polarisation in wavelength bins close to the
line core ($\lambda_2$, $\lambda_3$), suspecting significant influence 
of Faraday rotation on our inference. For the various correlations computed 
from these wavelength bins, both $EB$ and $TE$ correlations show a sign
reversal, while the $TB$ correlations do not. This indicates that the sign
of $E$ changes (and $B$ does not change) closer to the line core. We
mostly saw this reversal in the sign of $E$ at $\lambda_2$, except for
one AR where it happened at $\lambda_3$. However, this simply depends
on the Doppler shift of the spectral line. We performed tests with
a simple model of the solar atmosphere, and different iterations of it,
to investigate the cause of this sign reversal of $E$ and to reproduce it. 
Finally, we performed inversions of the observed profiles by HMI to infer
the atmospheric parameters. From these computed synthetic spectra with and without
Faraday rotation, we observed in both cases the sign reversal of $E$ 
at $\lambda_2$, thus ruling out Faraday rotation as the cause of the sign reversal.
It is still unclear what exactly causes this sign reversal of
$E$ near the line core, which occurs higher up in the atmosphere.
$E$ polarisation is linked to the topology of the magnetic field. Therefore,
to understand this better, synthesising spectra from three-dimensional MHD
simulations might be required to capture the changing magnetic
field topology with height and the radiative transfer effects fully.
This would help us narrow down the relation between the sign of $E$
and the topology of the magnetic field, while still accounting for
magneto-optic effects, mainly Faraday-rotation.
The tests also revealed that the contributions
to $c_{\rm A}^{EB}$ purely from Faraday rotation are relatively insignificant
away from the nominal line core (Fig.~\ref{syn_fararot}). This agrees 
with the conclusions 
of \cite{Bran_eb_2} who found that, provided the contributions 
from Faraday rotation are subdominant compared with the helicity contributions, 
one can detect magnetic helicity by using the $EB$ decomposition. In the solar 
context, this is true away from the core of the spectral line. Therefore, we
can safely infer magnetic helicity employing the $EB$ decomposition.
On the other hand, the spatial pattern of $B$ polarisation also has
an interesting feature in that it is predominantly bipolar (see Fig.~\ref{EB_maps}) 
for almost all the inspected ARs at all wavelength bins. This is important
since any spatial smoothing or averaging, even after multiplying with $E$
or any other parity-even quantity, will result in cancellation.

The formalism to obtain $E$ and $B$ polarisation
relies on linear polarisation, as it is directly
borrowed from cosmology, wherein Thompson scattering only generates linear 
polarisation. However, in the solar context, the most frequently used diagnostic
is the Zeeman effect, which also generates circular polarisation, and hence
Stokes $V$ is non-zero. Stokes $V$ carries with it additional information
about the directionality of the line-of-sight magnetic field.
Except for the cases where we inferred Stokes $Q$ and $U$ from the
components of the transverse magnetic field through Eq.~(\ref{PfromB}),
Stokes $V$ has not been used in the present study.
Including it is another possible next step to extend
the present formalism to invoke Stokes $V$ together with the $EB$ decomposition.

\begin{acknowledgements} 
We thank Nishant Singh (Pune, India) for useful discussions
during the early phase of this project.
Support through the NSF Astrophysics and Astronomy Grant Program,
grant 1615100, and the Swedish Research Council, grant 2019-04234 (AB)
are gratefully acknowledged.
MJK acknowledges the support of 
the Academy of Finland ReSoLVE Centre of Excellence (grant No.~307411).
AP was funded by the International Max Planck Research School 
for Solar System Science at the University of G\"ottingen.
This project has received funding from the European Research Council
under the European Union's Horizon 2020 research and innovation 
programme (project "UniSDyn", grant agreement n:o 818665).
\end{acknowledgements}

   \bibliographystyle{aa} 
   \bibliography{lit}

\begin{appendix}

\section{$E$ and $B$ decomposition}
\label{app1}
To study the polarisation signals of the cosmic microwave background (CMB),
the linear polarisation signals generated through Thomson scattering are
decomposed into $E$ and $B$ polarisations \citep{Kamionkowski_1997,Zaldarriaga}. 
To demonstrate this decomposition here, we follow the convention and 
approach of \cite{Zaldarriaga}, which arose out of the need to extract power 
spectra based on the rotationally invariant linear polarisation parameters. 
For a detailed derivation, we refer to the original article and the references therein.
Here, we focus on the small-scale limit and discuss different conventions.

Stokes $Q$ and $U$ are frame-dependent
quantities: a rotation of the polarisation basis ($\hat{\vec{e}}_1$,$
\hat{\vec{e}}_2$) by an angle $\phi$ in the 
plane perpendicular to the propagation direction $\hat{\vec{n}}$ transforms
$Q$ and $U$ as
\begin{align}
  (Q\pm iU)' = e^{\mp 2i\phi}(Q\pm iU) (\hat{\vec{n}}),
\end{align}
with $\hat{\vec{e}}_1^{'} = \cos{\phi}~\hat{\vec{e}}_1 +\sin{\phi}
~\hat{\vec{e}}_2$ and $\hat{\vec{e}}_2^{'} = -\sin{\phi}~\hat{\vec{e}}_1 +
\cos{\phi}~\hat{\vec{e}}_2$. For a harmonic analysis of the $Q+iU$ over the
entire sphere and given the rotational dependence of $Q$ and $U$,
it is appropriate to expand them in a spin-weighted basis as
\begin{align}
  (Q\pm iU) (\hat{\vec{n}}) &= \sum_{lm} a_{lm}^{\pm 2}~ _{\pm 2}Y_{lm}(\hat{\vec{n}}),
   \label{spin_up_dn}
\end{align}
where $_s Y_lm$ are spin-weighted spherical harmonic functions for each 
integer $s$ with $|s| \leq l$, which transform under rotation. For convenience, 
we can define linear combinations of the above coefficients,
such as
\begin{equation}
  a_{lm}^E = -(a_{lm}^2+ a_{lm}^{-2})/2 \quad\mbox{and}\quad
  a_{lm}^B = -(a_{lm}^2- a_{lm}^{-2})/2i.\label{lc_zs}
\end{equation}
Here one can also notice the parity-even and parity-odd 
properties of $E$ and $B$; $E$ remains unchanged, whereas $B$ changes sign.

In this paper, we work within the confines of the small-scale limit.
That is, for a high enough degree of spherical harmonics, we can
neglect the curvature of the sphere and consider it as a plane 
normal to $\vec{e}_z$. In this limit, spin-weighted spherical harmonics
can be approximated in terms of exponentials as
\begin{equation}
\begin{split}
 _2 Y_{lm} = \left[\frac{(l-2)!}{(l+2)!}\right]^{1/2}\dcut^2 Y_{lm}\longrightarrow 
 \frac{1}{2\pi} \frac{1}{l^2}~ \dcut^2 ~ e^{i\vec{k}\cdot\vec{x}}, \label{ssl}\\
  _{-2} Y_{lm} = \left[\frac{(l-2)!}{(l+2)!}\right]^{1/2}\overline{\dcut}^2 Y_{lm}\longrightarrow 
  \frac{1}{2\pi} \frac{1}{l^2}~ \overline{\dcut}^2 ~ e^{i\vec{k}\cdot\vec{x}},
\end{split}
\end{equation}
where $\vec{x}$ is a vector in the plane normal to $\vec{e}_z$ and $\vec{k}$ is 
its counterpart in Fourier-space, where $k_x+ik_y = ke^{i\phi_k}$. Furthermore, $\dcut$ and  
$\overline{\dcut}$ are spin raising and spin lowering operators (see \citealt{goldberg}).

Thus, invoking the small-scale approximation (\ref{ssl})
and using the linear combinations defined in 
Eq.~(\ref{lc_zs}), we can obtain the following expression from Eq.~(\ref{spin_up_dn})
(see \citealt{Zaldarriaga})
\begin{equation}
  \tilde{Q}+i\tilde{U} = (\tilde{E}+i\tilde{B}) ~ e^{2i\phi_k} \label{ssl_1}.
\end{equation}
The relation can also be written differently in terms of the
components of the unit vector $\hat{\vec{k}} = \vec{k}/k$ as
\begin{equation}
 (\tilde{E}+i\tilde{B}) = (\hat{k}_x - i\hat{k}_y)^2 (\tilde{Q}+i\tilde{U}), \label{ssl_2}
\end{equation}
which is the relation used in the present paper.

  \begin{figure}
    \centering
    \includegraphics{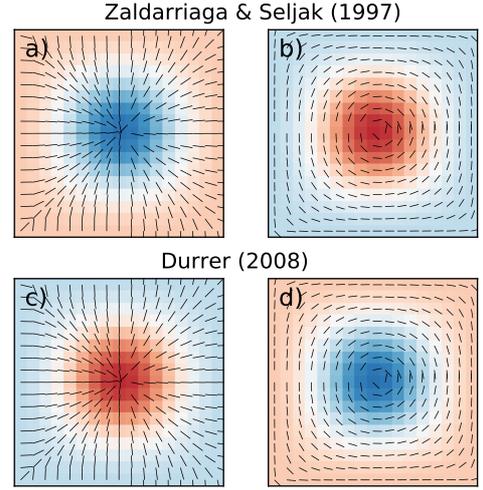}
    \caption{Top row: illustration of the pattern of polarisation generated by positive (red) and
    negative (blue) values of $E$ following the sign convention of \cite{Zaldarriaga}. 
    Bottom row: same as in top row, but for the sign convention of \cite{Durrer}.}
    \label{conv}
\end{figure}

The linear combinations in Eq.~(\ref{lc_zs}) were defined according to 
the convention chosen by \cite{Zaldarriaga} and this is the sign convention we follow in this
study. However, there exists another sign convention followed by \cite{Durrer} 
and \citet{Bran_eb_2}, wherein these linear combinations are defined as
\begin{equation}
  \tilde{a}_{lm}^E = (a_{lm}^2+ a_{lm}^{-2})/2 \quad\mbox{and}\quad
  \tilde{a}_{lm}^B = (a_{lm}^2- a_{lm}^{-2})/2i.\label{lc_d}
\end{equation}

As a result of this convention, Eq.~({\ref{ssl_2}}) 
acquires an additional minus sign. The sign convention chosen by \cite{Zaldarriaga}
is such that positive (negative) values of $E$ generate a tangential (radial) pattern (see Figs.~\ref{conv}a,b)
and in the case of $B$ polarisation, positive (negative) values of $B$ generate an anticlockwise (clockwise) 
inward spiralling pattern of polarisation.
A consequence of the different sign convention of \cite{Durrer} is that negative values of $E$
now generate a tangential pattern of polarisation (see Figs.~\ref{conv}c, d).

\section{Examples}
\label{app2}

We performed tests with a magnetic field
configuration following Sect.~2.3 of \cite{Bran_eb_1}. The magnetic
field was defined as a sum of gradient- and curl-type fields:
\begin{align}
  &\vec{b}(x,y) = \vec{F} + \vec{G}, \label{syn1}\\
  & F_i (x,y) = \partial_i f, \quad G_i(x,y) = \epsilon_{ij} \partial_j g,
  \\
  & f = f_0 \cos(kx) \cos(ky), \quad g = g_0 \cos(kx)\cos(ky). 
  \label{syn3}
\end{align} 

\begin{figure}
    \centering
    \includegraphics{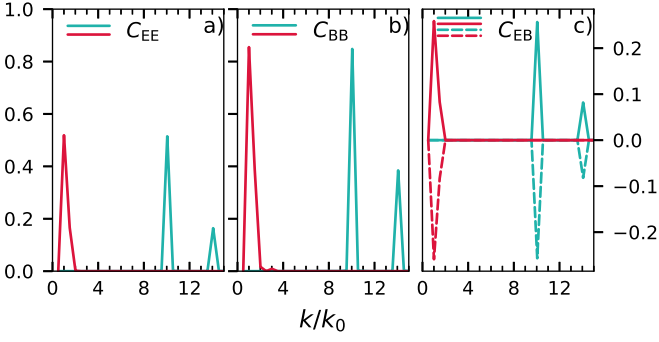}
    \caption{$c_{\rm A}^{EE} (k)$, $c_{\rm A}^{BB} (k)$, and $c_{\rm A}^{EB} 
(k)$ calculated from Stokes $Q$ and $U$ for case a: pure $E$ polarisation, 
case b: pure $B$ polarisation, and case c: non-zero $E$ and $B$ polarisations,
respectively. The red curve is for $k=k_0$ and the blue curve for $k=10k_0$. 
The dotted curve in panel (c), corresponds to a different handedness of the 
original $\vec{b}$ field resulting in $c_{\rm A}^{EB} (k)$ of an opposite 
sign.}
    \label{syn_simeg}    
\end{figure}

This $\vec{b}$ vector only provides the planar projection of
a fully three-dimensional solenoidal magnetic field.
With this, we have the freedom to choose a vector field with a given 
wavenumber $k$. With such a field in a simple atmosphere, we can
synthesise the full Stokes vector and compute $E$ and $B$ from Stokes $Q$ 
and $U$. The goal is to exploit the $E$ and $B$ decomposition of
linear polarisation to infer the characteristics of the original vector field 
(be it the wavenumber or the handedness of the vector field) directly from 
the polarisation signal.
We chose three cases: one with a $\vec{b}$ field corresponding to a pure $E$ 
polarisation ($f_0 = 1, g_0=0$, Fig.~\ref{syn_simeg}a), a second 
case with a $\vec{b}$ field corresponding to pure $B$ polarisation ($f_0 = 
1, g_0=1$, Fig.~\ref{syn_simeg}b), and lastly, a $\vec{b}$ field, which 
would result in both $E$ and $B$ polarisations ($f_0=\cos\theta, g_0 = \pm
\sin\theta$, Fig.~\ref{syn_simeg}c) but of opposite handednesses by
changing the sign of $g_0$. In all three cases, we repeated the 
experiments for different wavenumbers, $k=k_0$ and $k=10k_0$. As described 
before, we synthesised Stokes $Q$ and $U$ for all these cases and computed 
the relevant shell-integrated spectra. We show $c^{EE} (k)$ for the case of 
pure $E$ polarisation, $c^{BB} (k)$ for pure $B$ polarisation, and $c^{EB} 
(k)$ for 
the third case where both $E$ and $B$ are non-zero.

In all cases, we retrieve maximum amplitudes in the corresponding 
normalised correlation spectra at the chosen wavenumbers
to define the $\vec{b}$ fields with $k=k_0$ and $k=10k_0$. For the 
last case (Fig.~\ref{syn_simeg}c) we also retrieve different signs 
of $c^{EB} (k)$ for the opposite handednesses of the vector fields. This is 
due to the $B$ polarisation changing sign under a parity transformation.
For $k=10k_0$, we also retrieve a secondary peak of lower amplitude. 
This is probably an artifact resulting from the spectral synthesis, wherein 
we assume a simplified atmosphere with these idealised magnetic fields.

\end{appendix}
\end{document}